\documentclass[conference]{IEEEtran}
\IEEEoverridecommandlockouts
\usepackage{cite}
\usepackage{amsmath,amssymb,amsfonts}
\usepackage{algorithmic}
\usepackage{graphicx}
\usepackage{textcomp}
\usepackage{xcolor}
\usepackage{hyperref}
\usepackage{float}
\usepackage{caption}
\usepackage{stfloats}
\usepackage{tikz}
\usepackage{pgfplots}
\pgfplotsset{compat=1.17}
\usepackage[a4paper, margin=1in]{geometry}
\usepackage{placeins}

\def\BibTeX{{\rm B\kern-.05em{\sc i\kern-.025em b}\kern-.08em
    T\kern-.1667em\lower.7ex\hbox{E}\kern-.125emX}}
\begin{document}

\title{LLM Based Web Accessibility Repair: An Empirical Study of Detection, Remediation, and Cost}

\author{
Oluwatoyosi Oyelayo, Ghada Abushaqra, Parham Asadi, Durjoy Dey, Diego Elias Costa\\
\textit{Department of Computer Science and Software Engineering}\\
\textit{Concordia University, Montreal, Canada}\\
\small
\begin{tabular}{c}
\texttt{oluwatoyosi.oyelayo@mail.concordia.ca} \\
\texttt{ghada.a.shaqra89@gmail.com} \\
\texttt{parham.5579@gmail.com} \\
\texttt{durjoy.dey@mail.concordia.ca} \\
\texttt{diego.costa@concordia.ca}
\end{tabular}
}

\maketitle

\section{\textbf{Abstract}}
Ensuring web accessibility at scale remains a significant challenge, as rule-based tools provide limited coverage while manual remediation is costly and error-prone. This paper investigates the effectiveness of large language models (LLMs), specifically the Kimi K2.5 model, for automated accessibility detection and repair in comparison to traditional rule-based approaches. For detection, LLMs achieve performance comparable to rule-based tools (F1 $\approx$ 0.65), with strong semantic understanding (F1 = 0.83) but lower reliability in syntactic (0.62) and layout-related (0.48) violations. For remediation, LLM-generated fixes are syntactically valid in over 99.7\% of cases and improve accessibility compliance in 80.2\% of instances, reducing violations from 3.98 to 1.7 per-file. However, fewer than 26\% of cases are fully resolved, and approximately 30\% of patches introduce structural changes.

We further evaluate efficiency and find that iterative agent-based refinement increases computational cost by 52\% and API usage by 1.64$\times$ without improving remediation outcomes. 
These findings indicate that while LLMs are effective for partial accessibility repair, they are insufficient for complete and reliable remediation. We conclude that scalable accessibility solutions require hybrid approaches that combine LLM capabilities with rule-based validation and constraint-aware correction mechanisms.

\begin{IEEEkeywords}
Web Accessibility, Large Language Models, Program Repair, WCAG Compliance, Cost Efficiency
\end{IEEEkeywords}

\section{\textbf{Introduction}}

Web accessibility remains a critical requirement for inclusive digital systems, ensuring that web content is perceivable, operable, understandable, and robust for users with disabilities. Despite the availability of standards such as WCAG 2.2 and automated accessibility checkers (e.g., axe-core and Lighthouse), a large proportion of websites still contain accessibility violations, particularly in dynamic interfaces, forms, and multimedia content. Traditional automated tools primarily detect rule-based violations but provide limited support for repairing issues and lack semantic understanding of interface intent \cite{b5}\cite{b6}.

Recent advances in large language models (LLMs) have enabled new approaches for automated accessibility detection, validation, and repair.\cite{b11} \cite{b12} \cite{b13} Several studies have explored the use of LLMs to analyze HTML structures and generate accessibility fixes. For example, AccessGuru applies LLM-based transformations to HTML elements and evaluates repairs on a large benchmark of real-world accessibility issues, reporting substantial reductions in violation scores compared with baseline methods \cite{b1}. Similarly, generative approaches have been applied to specific accessibility domains such as web forms and STEM images, where LLM-based tools generate labels, descriptions, or alternative text to reduce accessibility barriers \cite{b2}\cite{b3}. Other works investigate LLM-assisted auditing and validation, showing that language models can complement rule-based tools by interpreting accessibility requirements and refining automated assessments \cite{b5}\cite{b6}. Multi-agent accessibility repair frameworks further extend this paradigm by coordinating specialized agents (e.g., planner, fixer, validator) and preserving functional correctness during automated remediation in modern web applications \cite{b4}.

However, existing LLM-based accessibility repair and validation studies share several methodological limitations. First, most approaches evaluate repair success primarily using automated accessibility checkers, which verify rule-based WCAG compliance but do not assess semantic adequacy, visual integrity, or functional correctness of the repaired interface \cite{b1}\cite{b3}. Second, prior work rarely analyzes cost–performance trade-offs of LLM-based remediation, including token consumption and iterative validation overhead \cite{b1}\cite{b4}. Third, evaluation settings vary widely in scope and datasets, ranging from static HTML snippets and domain-specific components to limited real-world pages, making results difficult to compare across studies \cite{b1}\cite{b2}\cite{b3}. Consequently, the reliability and generalizability of LLM-generated accessibility repairs in realistic web environments remain insufficiently understood.

To address these gaps, this research investigates the design and evaluation of LLM-based agents (utilizing the Kimi K2.5 model) for automated web accessibility remediation with integrated validation mechanisms. Specifically, the study examines how effectively LLM agents detect and repair accessibility violations while preserving visual and structural correctness, evaluates the role of an additional custom validation layer in improving reliability, and analyzes the trade-off between token cost and remediation success under single-pass versus iterative strategies. Existing approaches typically validate LLM-generated fixes using a single automated accessibility checker, considering remediation successful when reported WCAG violations decrease \cite{b1}\cite{b3}. In contrast, this work introduces a multi-layer validation framework in which LLM-generated repairs are first assessed by an accessibility checker and subsequently verified through semantic and structural validation stages before acceptance (LLM $\rightarrow$ checker $\rightarrow$ semantic validator $\rightarrow$ structural validator $\rightarrow$ accept/reject). By establishing a unified evaluation framework that jointly measures accessibility compliance, interface integrity, and cost efficiency, this work aims to advance reliable and scalable autonomous accessibility remediation for modern web systems.

\section{Related Work}
LLMs are increasingly used to improve web accessibility by detecting WCAG violations, generating repairs, and validating accessibility beyond traditional rule-based tools. They can understand HTML semantics and assist in auditing complex accessibility cases that require human judgment. Recent research also explores combining LLM analysis with user behavioral signals to capture real usability barriers and enable more comprehensive accessibility evaluation. 

\textbf{1) LLM for accessibility detection/repair.}
Recent research has explored the use of large language models (LLMs) to automatically detect and repair web accessibility violations by leveraging their ability to interpret HTML structure and accessibility semantics. Early work demonstrates that LLMs can directly modify markup to correct WCAG violations. For instance, AccessGuru applies LLM-based transformations to HTML elements and evaluates repairs on a large benchmark of real-world accessibility issues using an automated accessibility checker to measure violation scores before and after repair \cite{b1}. The study reports up to (84\%) reduction in violation scores compared with lower baseline performance, establishing LLMs as effective automated remediation agents capable of generating accessibility-compliant attributes and structural corrections.
Beyond general HTML repair, several studies investigate domain-specific accessibility generation tasks. In web forms and interactive components, LLM-based approaches improve labeling, input associations, and feedback messages, with evaluation typically based on WCAG compliance checks and usability assessment of generated content \cite{b2}. Similarly, LLM-based accessibility tools have been applied to STEM images, where models generate descriptive alternative text for scientific diagrams and visualizations [3]. These approaches are evaluated through human judgment studies measuring semantic accuracy, completeness, and usefulness for visually impaired users, with results showing superior accessibility quality compared to traditional captioning tools \cite{b3}.
More recent work extends single-step repair toward structured remediation frameworks. Multi-agent accessibility systems coordinate specialized roles such as planning, repair, and validation to improve reliability and preserve interface functionality during automated corrections \cite{b4}. Evaluation in such systems focuses on repair correctness and functional preservation within controlled web application environments, typically through case-based validation rather than large-scale quantitative benchmarks\cite{b4}.
Despite promising results, existing LLM-based accessibility detection and repair studies remain limited in scope and evaluation methodology\cite{b8} \cite{b9}. Many approaches operate on static HTML fragments or domain-specific artifacts rather than full interactive web pages, and repair effectiveness is primarily measured through violation reduction or qualitative usability assessment \cite{b1}\cite{b3}. Consequently, the ability of LLM-generated repairs to preserve visual layout, structural integrity, and functional behavior across realistic web environments remains insufficiently explored \cite{b4}. These limitations motivate the need for more comprehensive evaluation and validation frameworks for LLM-based accessibility remediation, which this work addresses.\cite{b14}

\textbf{2) LLM for auditing/validation. }
Recent work has explored the use of large language models (LLMs) to support accessibility, auditing, and validation tasks beyond rule-based automated checkers. Traditional tools often fail to assess semantic accessibility aspects such as logical content structure or context-dependent WCAG criteria. To address this limitation, LLMs have been integrated into validation workflows to interpret accessibility techniques and refine automated tool outputs. For example, Garcia et al. \cite{b5} evaluate GPT-based prompting strategies for validating accessibility cases that require human judgment, showing that LLMs can assist in assessing complex WCAG techniques. Similarly, Chen et al. \cite{b6} compare LLM-based auditing with traditional tools and expert evaluation, demonstrating that language models can detect accessibility issues beyond automated checker coverage while still facing consistency limitations. These studies highlight the potential of LLMs as semantic accessibility validators, although evaluation remains largely qualitative and lacks standardized quantitative benchmarks.

\textbf{3) Behavioral/usage signals}
In addition to structural accessibility analysis, some research has explored the use of behavioral and usage signals to understand accessibility and usability barriers in web systems. User interaction data such as navigation paths, session duration, click depth, and task completion behavior can provide indirect evidence of accessibility issues that are not detectable through static code inspection alone. For example, the study by Garc{\'i}a et al. \cite{b7} analyzes web access logs using machine learning feature selection techniques to identify patterns associated with user interaction performance. Such behavioral features can reveal difficulties in navigation, content discovery, or form interaction that may indicate underlying accessibility or usability barriers. While our study focuses strictly on static structural analysis, future hybrid systems could incorporate these behavioral signals to validate LLM repairs against actual user experiences and interaction barriers.

\section{Research Contribution \& Differentiation}
This work makes three contributions:

(1) A comparative evaluation of LLM-based accessibility analysis against a rule-based baseline (Axe-Core) across detection, remediation, and cost.

(2) A multi-layer validation framework that evaluates fixes based on syntactic validity, structural preservation, and WCAG compliance.

(3) An empirical analysis demonstrating that iterative agent-based remediation increases computational cost without improving repair effectiveness.

\section{Research Questions}

To evaluate LLM-based accessibility analysis, this study addresses the following questions:

\begin{itemize}
    \item RQ1: How accurately can an LLM detect accessibility violations compared to Axe-Core?
    \item RQ2: How effectively can an LLM repair accessibility violations while preserving structure?
    \item RQ3: What is the cost of agent-based repair compared to zero-shot?
\end{itemize}

\section{Methodology}

\subsection{Problem Formulation}
This study investigates the effectiveness of Large Language Model (LLM)-based agents for web accessibility detection and remediation within a unified evaluation framework. The problem is decomposed into three interconnected computational tasks aligned with the research questions.

\textit{\textbf{First}}, accessibility detection is formulated as a binary classification problem. Let \(x_i \in X\) denote an HTML page and \(y_i \in \{0,1\}\) denote its ground truth label, where \(y_i = 1\) indicates the presence of accessibility violations and \(y_i = 0\) indicates a clean page. Each system \(f\) produces a prediction:
\begin{equation}
\hat{y}_i = f(x_i)
\end{equation}

\textbf{\textit{Second}}, accessibility remediation is modeled as an automated program repair problem, in which a transformation function g generates a modified version of the input HTML:

\begin{equation}
x_i' = g(x_i)
\end{equation}

The goal of this transformation is to reduce accessibility violations while preserving the structural and semantic integrity of the original page.

\textbf{\textit{Third}}, the efficiency of LLM-based remediation is analyzed as a cost-performance trade-off problem, where the objective is to balance repair quality against computational cost, measured in token usage and interaction complexity.

\subsection{System Architecture}
The system is designed as a dual-pipeline architecture consisting of a deterministic rule-based pipeline (Axe-Core) and a probabilistic reasoning-based pipeline (LLM agent), followed by a unified evaluation layer.\cite{b15} \cite{b16}
At a high level, each HTML page is processed independently by both systems. The rule-based pipeline applies predefined accessibility rules, while the LLM pipeline performs contextual reasoning over the HTML structure. The outputs of both systems are then evaluated using common metrics and analysis procedures.
This architecture can be conceptualized as a layered system:

Dataset → Data Loader → Two Engines → Violations → Evaluation

\begin{figure*}[t]
\centering
\includegraphics[width=\textwidth]{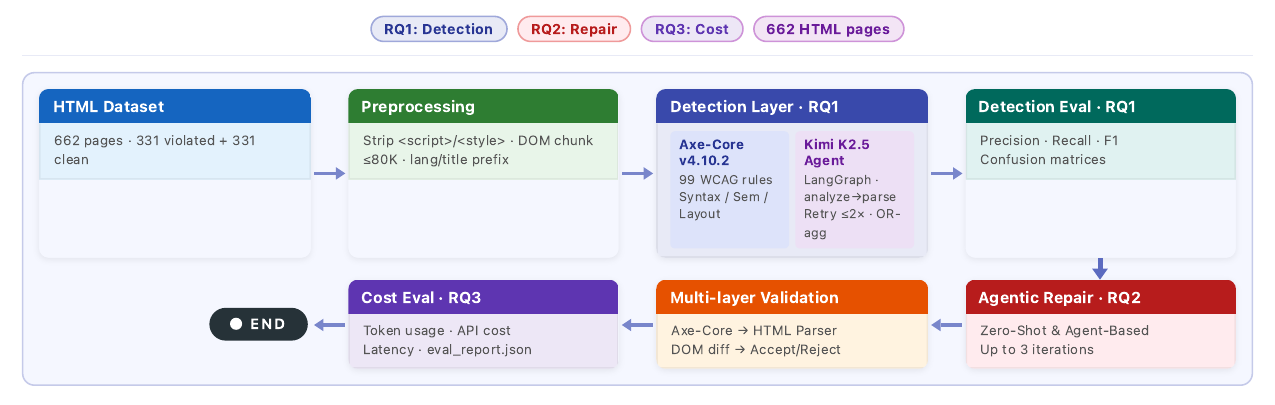}
\caption{Overview of the dual-pipeline architecture combining rule-based and LLM-based accessibility analysis with multi-layer validation.}
\label{fig:architecture}
\end{figure*}

\begin{itemize}
    \item Data Layer (HTML input): The data layer represents the input space of the system, consisting of raw HTML documents collected from both violated and corrected web pages. Each document serves as an independent sample in the evaluation process. This layer defines the problem domain by providing structured web content that reflects real-world accessibility scenarios, including forms, navigation components, and multimedia elements.
    \item Preprocessing Layer (cleaning and chunking): The preprocessing layer transforms raw HTML into a model-ready format. First, non-essential elements such as \texttt{<script>} and \texttt{<style>} tags are removed to reduce noise. Then, each document is segmented into smaller chunks to satisfy LLM input constraints while preserving local structural context.
    \item Detection Layer (Axe-Core and LLM agent): The detection layer applies two independent approaches to identify accessibility violations. The Axe-Core component uses deterministic rule-based analysis to detect violations according to predefined WCAG rules. Next, the LLM agent analyzes the HTML using contextual reasoning and structured prompting to identify both explicit and implicit accessibility issues.

    \item Agentic Repair Layer (LLM-based remediation): The agentic repair layer generates corrected HTML for pages containing violations. Unlike single-pass generation, this layer operates in an iterative loop where the LLM analyzes the input, proposes fixes, and refines them based on feedback. This process mimics human debugging by enabling incremental improvement.

    \item Validation Layer (multi-criteria verification): The validation layer ensures that generated fixes are correct, safe, and meaningful. Each repair is evaluated across three criteria: reduction of accessibility violations, syntactic correctness of HTML, and preservation of structural integrity.

    \item Evaluation Layer (metrics and analysis): The evaluation layer measures system performance across detection accuracy, remediation quality, and computational efficiency. Metrics such as precision, recall, F1 score, and confusion matrices are used for detection, while remediation is evaluated through compliance improvement and structural correctness. Cost metrics capture token usage and communication overhead.
\end{itemize}
This layered design ensures modularity, reproducibility, and clear separation of responsibilities across system components.

\subsection{Data and Preprocessing Layer}
The system is evaluated on a balanced dataset of 662 HTML pages, comprising 331 pages with accessibility violations and 331 corresponding corrected pages. This balanced design ensures an unbiased assessment of classification performance. All original samples contain violations, while the corrected samples are treated as improved reference versions. However, residual violations may still remain due to limitations in automated remediation and validation.

\begin{table}[h]
\centering
\caption{Violation Analysis on 662 Samples Before and After Fixing}
\label{tab:violation_analysis}
\renewcommand{\arraystretch}{1.2}
\begin{tabular}{lccc}
\hline
\textbf{Source} & \textbf{No Violation} & \textbf{Violation} & \textbf{Total} \\
\hline
scraped\_sites        & 0   & 331 & 331 \\
scraped\_sites\_fixed & 331 & 0   & 331 \\
\hline
\textbf{Total}        & 331 & 331 & 662 \\
\hline
\end{tabular}
\end{table}

Each HTML document undergoes preprocessing before analysis. Let $x_i$ denote the original HTML page. A cleaning function is applied:

\begin{equation}
x_i' = \operatorname{clean}(x_i)
\end{equation}

This step removes non-essential elements such as 
\texttt{<script>} and \texttt{<style>} tags to reduce noise and improve model focus.
Due to input size constraints of LLMs, each cleaned document is segmented into smaller units using a chunking function:
\begin{equation}
x_i' \rightarrow \{c_1, c_2, \ldots, c_k\}
\end{equation}
where each chunk $c_j$ satisfies the model’s token limit. Chunking ensures that large documents can be processed efficiently while preserving local structural context.

\subsection{Detection Layer}
The detection layer consists of two independent systems: a rule-based engine and an LLM-based agent.
 \subsubsection{Rule-Based Detection (Axe-Core)}
The rule-based system is defined as a deterministic function:
\begin{equation}
f_{axe}(x) = \text{rules}(x)
\end{equation}

Axe-Core (v4.10.2) evaluates each HTML page using a predefined set of accessibility rules derived from WCAG standards. The process involves rendering the page in a browser environment using Playwright, injecting the Axe-Core library, and executing the \texttt{axe.run()} function to detect violations.
This approach provides high reliability and consistency, as results are deterministic and reproducible.

\subsubsection{LLM-Based Detection}
The LLM agent is modeled as a probabilistic function:
\begin{equation}
f_{\text{llm}}(x) = \mathrm{LLM}(\text{prompt}, x)
\end{equation}
Each chunk $c_j$ is independently analyzed:
\begin{equation}
\hat{y}_j = f_{\text{llm}}(c_j)
\end{equation}

The final prediction for the page is obtained through aggregation:
\begin{equation}
\hat{y} = \max_{j} \left(\hat{y}_j\right)
\end{equation}

This aggregation ensures that if any portion of the page contains a violation, the entire page is classified as containing violations.
The use of an LLM enables contextual reasoning, allowing the system to detect issues that are not explicitly defined by rules.

\subsection{Violation Taxonomy}
To support fine-grained analysis, accessibility violations are categorized into three types: syntactic, semantic, and layout violations.
Syntactic violations correspond to structural errors in HTML, such as missing required attributes or invalid nesting. These violations are deterministic and can be verified using rule-based methods.
Semantic violations relate to the meaning and interpretation of elements, particularly for assistive technologies. These include incorrect ARIA roles, missing labels, and misuse of semantic elements, and require contextual understanding.
Layout violations affect visual accessibility and interaction, including issues such as color contrast and navigation structure. These violations often require both measurable criteria and higher-level reasoning.
This taxonomy enables detailed comparison of system performance across different dimensions of accessibility.

\subsection{Agentic Remediation Layer}

Accessibility remediation is modeled as an iterative transformation process:
\begin{equation}
x^{t+1} = g(x^t)
\end{equation}

where $x^t$ represents the HTML at iteration t. The LLM agent operates in a loop consisting of analysis, repair, and refinement steps.
Unlike zero-shot prompting, which produces a single output, the agentic approach allows for iterative improvement, enabling the system to correct its own errors and refine its outputs over multiple steps.
This design reflects real-world debugging workflows and improves reliability in code generation.

\subsection{Multi-Layer Validation Framework}

A key contribution of this work is the introduction of a multi-layer validation framework that evaluates generated repairs across multiple dimensions.
A repair $x^{\prime}$ is considered valid if it satisfies the following conditions:

\begin{equation}
\text{Fix} = 1 \iff
\begin{cases}
V(x') < V(x) \\
\text{Parser}(x') = \text{valid} \\
\text{Structure}(x') \approx \text{Structure}(x)
\end{cases}
\end{equation}

where V(x) denotes the number of accessibility violations detected by Axe-Core.
The first condition ensures that the repair improves WCAG compliance. The second condition verifies syntactic correctness using an HTML parser. The third condition ensures that the structural integrity of the document is preserved by comparing DOM trees before and after modification.
This multi-objective validation addresses a key limitation in prior work, where success is often defined solely by reduction in violation counts.

\subsection{Cost and Efficiency Analysis}
To evaluate practical feasibility, the study measures the computational cost of LLM-based remediation.
The total cost is defined as:

\begin{equation}
\text{Cost} = \sum_{i=1}^{N} (\text{Token}_i \times \text{Price})
\end{equation}

where $Tokens_i$ represents the number of tokens used in each interaction.
Communication overhead is defined as the number of interactions required between the agent and the LLM:
\begin{equation}
\text{Overhead} = \#\text{Interactions}
\end{equation}

These metrics enable analysis of the trade-offs between repair quality and resource consumption.

\subsection{Evaluation Metrics}
Detection performance is evaluated using precision, recall, and F1 score:

\begin{equation}
\text{Precision} = \frac{TP}{TP + FP}, \quad
\text{Recall} = \frac{TP}{TP + FN}
\end{equation}

\begin{equation}
F1 = 2 \cdot \frac{\text{Precision} \cdot \text{Recall}}{\text{Precision} + \text{Recall}}
\end{equation}

Additionally, confusion matrices are used to analyze classification behavior in terms of true positives, false positives, true negatives, and false negatives.
Remediation performance is evaluated based on WCAG improvement, syntactic validity, and structural preservation, while efficiency is measured in terms of token usage, cost, and communication overhead.

\section{Experimental Procedure}

The experiment is conducted through a unified pipeline that begins with constructing a balanced dataset of HTML pages containing accessibility violations and their corresponding corrected versions, along with ground truth labels. Each HTML file is preprocessed by removing non-relevant elements and, when necessary, partitioned into context-preserving chunks to enable efficient analysis. Two detection approaches are then executed: a rule-based method using Axe-Core within a headless browser environment, and an LLM agent running Kimi K2.5 that analyzes the HTML through a structured analyze–parse–retry workflow, with results aggregated at the file level. The outputs of both systems are integrated with the ground truth to evaluate detection performance using standard classification metrics.
For samples identified as containing violations, the experiment proceeds to a remediation stage where two strategies are applied: a zero-shot generation approach and an iterative agent-based refinement process. The resulting patched HTML is subsequently re-evaluated using Axe-Core to quantify compliance improvement. In parallel, the system records token usage, latency, and retry behavior to assess computational cost and efficiency. The entire process produces structured outputs, including detection results, remediation performance metrics, and cost analysis reports, enabling a comprehensive evaluation of both effectiveness and efficiency.

To answer the research questions, the project methodology is structured into three distinct computational problem domains:

\subsection{\textbf{Detection as Binary Classification (Addressing RQ1)}}

\textbf{Aim. } To characterize where each system performs better - not which is superior overall - by comparing Axe Core, a deterministic rule-based engine, against an LLM agent across syntactic, semantic, and layout violation types.

\textbf{Approach: } Detection is framed as a binary classification task. For each HTML snippet, each system independently determines whether a violation is present or absent. The evaluation combines HTML snippets with known violations drawn from the AccessGuru dataset alongside subsets corrected by authors with frontend expertise, serving as verified negative examples. Violations are pre-categorized into three types - syntactic (e.g., missing alt attributes), semantic (e.g., incorrect ARIA roles), and layout (e.g., broken tab order) - to enable fine-grained comparative analysis.

\textbf{Evaluation: } The evaluation is conducted on a balanced dataset of 662 web samples, consisting of equal numbers of violating and corrected instances. The rule-based baseline is implemented using a widely adopted accessibility engine, while the LLM-based agent operates through structured prompt-driven analysis.
The task is formulated at two levels:
\begin{itemize}
    \item \textbf{Binary classification}: determining whether a page contains any accessibility violation.
    \item \textbf{Category-level detection}: identifying violations within syntactic, semantic, and layout dimensions.
\end{itemize}
\subsubsection{Binary Detection Performance}
Table II presents the overall performance of both systems in detecting whether a page contains violations.
\begin{table}[h]
\caption{Performance Comparison of Systems}
\label{tab:performance}
\centering
\vspace{0.2cm}

\begin{tabular}{lccc}
\hline
\textbf{System} & \textbf{Precision} & \textbf{Recall} & \textbf{F1 Score} \\
\hline
Axe-Core  & 0.51 & 0.93 & 0.66 \\
LLM Agent & 0.50 & 0.94 & 0.65 \\
\hline
\end{tabular}
\end{table}

Both approaches achieve high recall $(>0.93)$, indicating strong sensitivity to the presence of violations. However, precision remains relatively low $(~0.50)$, reflecting a tendency to produce false positives.
From a system-level perspective, the results demonstrate that the LLM-based agent achieves performance parity with the rule-based engine in coarse-grained detection. This suggests that LLMs are capable of approximating traditional rule-based systems in identifying the presence of accessibility issues.

\subsubsection{Category-Level Detection Behavior}
To better understand system behavior, Table III reports detection rates and agreement across violation categories.

\begin{table}[h]
\caption{Comparison of Rule-Based and LLM-Based Approaches}
\label{tab:comparison}
\centering

\begin{tabular}{lccc}
\hline
\textbf{Category} & \textbf{Rule-Based (\%)} & \textbf{LLM-Based (\%)} & \textbf{Agreement (\%)} \\
\hline
Syntax   & 45 & 91 & 48 \\
Semantic & 82 & 85 & 71 \\
Layout   & 40 & 57 & 50 \\
\hline
\end{tabular}
\end{table}

The table highlights notable differences across categories. For syntax, the LLM detects 91\% of cases compared to 45\% by the rule-based system, with only 48\% agreement, indicating that the LLM is much more sensitive but likely over-detecting issues. For semantic violations, both systems show similar performance (85\% vs. 82\%) and a higher agreement of 71\%, suggesting consistent and reliable detection. In layout, the LLM detects 57\% compared to 40\% by the rule-based approach, but agreement drops to 50\%, reflecting differing interpretations of layout-related accessibility issues.

Fig.~2 compares the detection rates of Axe-Core and the proposed agent across different accessibility violation categories.

\begin{figure}[t]
\centering
\begin{tikzpicture}
\begin{axis}[
    ybar=6pt, 
    bar width=7pt,
    width=\columnwidth,
    height=5.5cm,
    ylabel={Detection Rate},
    xlabel={Category},
    symbolic x coords={Syntax, Semantic, Layout},
    xtick=data,
    ymin=0, ymax=1,
    grid=major,
    grid style={dashed,gray!20},
    legend style={
        at={(0.5,1.12)},
        anchor=south,
        legend columns=2,
        draw=none
    },
    enlarge x limits=0.3
]

\addplot[
    fill=blue!70,
    draw=blue!90,
    nodes near coords,
    every node near coord/.append style={
        font=\scriptsize,
        xshift=-3pt 
    }
] coordinates {
    (Syntax,0.45)
    (Semantic,0.82)
    (Layout,0.40)
};

\addplot[
    fill=orange!85,
    draw=orange!95,
    nodes near coords,
    every node near coord/.append style={
        font=\scriptsize,
        xshift=3pt 
    }
] coordinates {
    (Syntax,0.91)
    (Semantic,0.85)
    (Layout,0.57)
};

\legend{Axe-Core, Agent}

\end{axis}
\end{tikzpicture}
\caption{Detection rate comparison across categories.}
\label{fig:detection}
\end{figure}
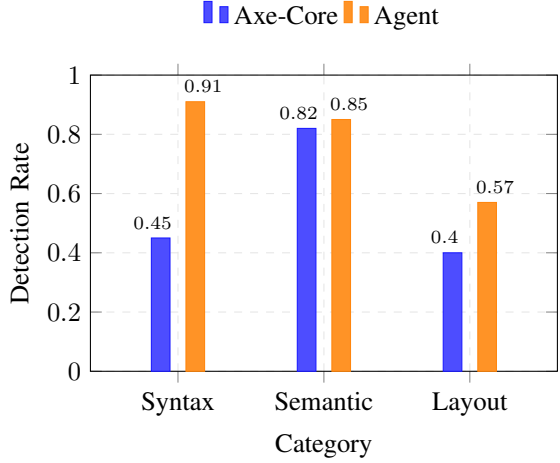
Overall, the agent demonstrates comparable detection performance to the rule-based system, with higher sensitivity in syntax and layout but lower precision, while maintaining strong semantic detection.

\subsubsection{Relative Performance Against Rule-Based Reference}
To quantify alignment, the LLM agent is evaluated against the rule-based engine as a reference. The results are summarized in Table IV.

\begin{table}[h]
\caption{LLM Performance Relative to Rule-Based Reference}
\label{tab:llm_performance}
\centering

\begin{tabular}{lccc}
\hline
\textbf{Category} & \textbf{Precision} & \textbf{Recall} & \textbf{F1-score} \\
\hline
Syntax   & 0.47 & 0.94 & 0.62 \\
Semantic & 0.82 & 0.84 & 0.83 \\
Layout   & 0.41 & 0.58 & 0.48 \\
\hline
\end{tabular}
\end{table}

The table evaluates how well the LLM aligns with the rule-based reference across categories using precision, recall, and F1-score.
Syntax: The LLM achieves very high recall (0.94) but low precision (0.47), resulting in an F1-score of 0.62. This indicates that the model captures almost all rule-based syntax violations but introduces many false positives, reflecting over-detection.
Semantic: This category shows the strongest performance, with balanced precision (0.82) and recall (0.84), leading to a high F1-score of 0.83. This suggests high reliability and strong agreement with the rule-based system.
Layout: Performance is weaker, with low precision (0.41) and moderate recall (0.58), yielding an F1-score of 0.48. This indicates limited alignment and inconsistency in detecting layout-related issues.
The LLM demonstrates high recall across all categories, but precision varies significantly, particularly in syntax and layout detection.

The results indicate that the LLM-based approach achieves comparable performance to the rule-based system at a high level, but its effectiveness varies significantly across categories. It demonstrates strong capability in semantic detection, with balanced precision (0.82) and recall (0.84), indicating reliable alignment with rule-based judgments. However, in syntactic detection, the LLM shows very high recall (0.94) but low precision (0.47), suggesting a tendency to over-report violations. For layout, both precision (0.41) and recall (0.58) are lower, reflecting weaker and less consistent performance. Overall, these findings suggest that while LLMs are particularly effective for context-dependent semantic analysis, they lack precision in rule-based scenarios and exhibit divergence in layout interpretation, supporting the need for a complementary hybrid approach.
\subsection{\textbf{Remediation as Automated Program Repair (Addressing RQ2)}}

\textbf{Aim. } To measure the ability of an LLM agent to autonomously generate accessible, structurally sound code patches for violations identified in RQ1.

\textbf{Approach: } The LLM agent receives HTML snippets containing known violations and is prompted to produce corrected code in an agentic loop. Fixes are generated across the full range of violation types established in RQ1. The same is done using zero-shot prompting and results are compared.

\textbf{Evaluation: } Each generated fix is assessed across three dimensions: syntactic validity via an HTML parser, WCAG compliance improvement via a follow-up Axe Core scan, and structural integrity via element-level inspection to confirm the original document structure is preserved. The results from the LLM agent are compared with the results from zero-shot prompting.

\subsubsection{Agent Iteration Effectiveness}

To better understand the impact of the agent’s iterative loop, we analyze how many iterations were required before validation success. 

\begin{figure}[t]
\centering
\begin{tikzpicture}
\begin{axis}[
    ybar,
    bar width=14pt,
    width=\columnwidth,
    height=5.5cm,
    ylabel={Number of Files},
    xlabel={Iterations},
    symbolic x coords={1, 2, 3 (max)},
    xtick=data,
    ymin=0,
    ymax=230,
    enlarge y limits={upper=0.15}, 
    grid=major,
    grid style={dashed,gray!20},
    enlarge x limits=0.35,
    nodes near coords,
    every node near coord/.append style={
        font=\footnotesize,
        yshift=3pt
    }
]

\addplot[fill=blue!70, draw=blue!90] coordinates {
    (1,208)
    (2,7)
    (3 (max),95)
};

\end{axis}
\end{tikzpicture}
\caption{Agent iteration effectiveness across iterations.}
\label{fig:agent_iterations}
\end{figure}
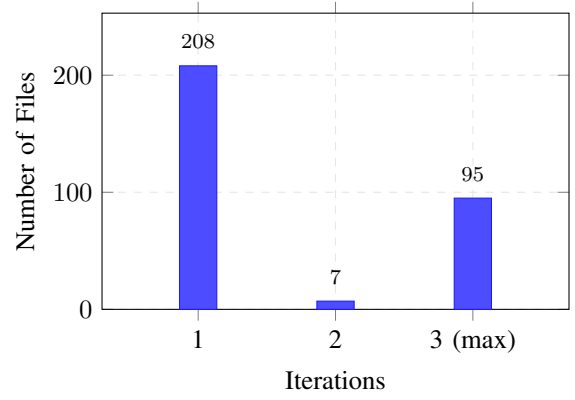

As shown in Fig. 3, 67\% of files passed validation on the first attempt, 2\% required exactly two iterations, and 31\% reached the maximum of three iterations without successfully resolving validation issues.
This distribution indicates that retries rarely contribute to successful remediation. When the first attempt fails, subsequent iterations are unlikely to correct the issue, suggesting that the agent does not effectively utilize validation feedback to improve its output.
These findings challenge a common assumption in agent-based systems: that iterative refinement leads to improved results. In this case, iterative self-correction does not enhance remediation quality and instead introduces additional computational overhead without measurable benefit.

\subsubsection{Overall Remediation Performance}
Table V presents the comparative performance of both approaches.

\begin{table}[h]
\caption{Remediation Performance Comparison}
\label{tab:remediation}
\centering

\begin{tabular}{lcc}
\hline
\textbf{Metric} & \textbf{Zero-Shot} & \textbf{Agent-Based} \\
\hline
Syntactic validity        & 99.7\% & 100\% \\
Structure preserved       & 68.2\% & 69.5\% \\
Compliance improved       & 80.2\% & 80.2\% \\
Fully fixed               & 25.7\% & 23.7\% \\
Avg violations (before)   & 3.98   & 3.98   \\
Avg violations (after)    & 1.68   & 1.71   \\
Avg violations reduced    & 2.44   & 2.42   \\
Avg iterations            & 1.0    & 1.64   \\
\hline
\end{tabular}
\end{table}

To provide a more comprehensive evaluation of remediation outcomes. Table VI provides a comprehensive summary of remediation effectiveness, including structural preservation and violation-level metrics.

\begin{table*}[t]
\centering
\caption{Summary of Accessibility Remediation Effectiveness}
\label{tab:strategy_comparison}

\footnotesize
\setlength{\tabcolsep}{2pt}

\resizebox{0.98\textwidth}{!}{
\begin{tabular}{lcccccccccc}
\hline
\textbf{strategy} &
\textbf{n\_files} &
\shortstack{\textbf{syntactic}\\\textbf{valid rate}} &
\shortstack{\textbf{structure}\\\textbf{preserved rate}} &
\shortstack{\textbf{avg structure}\\\textbf{similarity}} &
\shortstack{\textbf{compliance}\\\textbf{improved rate}} &
\shortstack{\textbf{fully}\\\textbf{fixed rate}} &
\shortstack{\textbf{avg violations}\\\textbf{before}} &
\shortstack{\textbf{avg violations}\\\textbf{after}} &
\shortstack{\textbf{avg violations}\\\textbf{reduced}} &
\shortstack{\textbf{avg}\\\textbf{iterations}} \\
\hline

agent & 308 & 1.00 & 0.69 & 0.83 & 0.80 & 0.24 & 3.98 & 1.71 & 2.42 & 1.64 \\
zero\_shot & 308 & 1.00 & 0.68 & 0.83 & 0.80 & 0.26 & 3.98 & 1.68 & 2.44 & 1.00 \\

\hline
\end{tabular}
}

\end{table*}

The results indicate that both zero-shot and agent-based approaches achieve nearly identical remediation performance across all metrics. Syntactic validity is effectively solved ($\geq 99.7\%$)
, while compliance improvement is limited to 80.2\% of cases and fewer than 26\% of files are fully fixed, highlighting the limitations of LLMs in achieving complete accessibility compliance. Structural preservation remains moderate, with approximately 30\% of patches altering the original layout. Notably, the agent-based approach requires more iterations (1.64 vs. 1.0) without delivering any measurable improvement, indicating that iterative refinement does not enhance remediation quality. This suggests that remediation success is largely determined in the initial generation, and that high-level validation feedback is insufficient to guide corrective behavior. The remaining violations likely involve complex, context-dependent issues that cannot be resolved from static HTML alone. From a practical perspective, zero-shot prompting emerges as the more efficient strategy, achieving equivalent outcomes without the additional computational overhead of agent-based iteration.

While RQ2 evaluates remediation effectiveness, RQ3 examines whether these improvements justify the associated computational cost.

\subsection{Cost and Efficiency Analysis (Addressing RQ3)}

\textbf{Aim.} To measure the trade-offs between cost, communication overhead, and remediation performance when using an LLM agent.

\textbf{Approach.} Building on RQ2, we measure the token cost and communication overhead involved in executing the LLM agent and compare them with those of zero-shot prompting.

\textbf{Evaluation.} We evaluate efficiency across API usage, token consumption, latency, and monetary cost.

\subsubsection{Aggregate Cost Analysis}

\begin{table}[H]
\caption{Cost and Resource Utilization}
\centering
\begin{tabular}{lccc}
\hline
Metric & Zero-Shot & Agent-Based & Ratio \\
\hline
API calls & 308 & 504 & 1.64$\times$ \\
Prompt tokens & 8.0M & 11.5M & 1.44$\times$ \\
Completion tokens & 6.2M & 9.6M & 1.55$\times$ \\
Total tokens & 14.2M & 21.2M & 1.49$\times$ \\
Total cost (USD) & \$4.92 & \$7.50 & $1.52\times$ \\
Avg latency per call & 474 ms & 433 ms & $0.91\times$ \\
Retries & 0 & 196 & -- \\
\hline
\end{tabular}
\end{table}
Table VII shows that the agent-based approach introduces substantial overhead across all cost dimensions. It requires 1.64× more API calls and 1.49× more tokens, resulting in a 52\% increase in total cost. This overhead is primarily driven by 196 retry operations, which expand both prompt and completion tokens due to repeated interaction cycles. Despite the increased cost, the average latency per call remains comparable, indicating that inefficiency arises from repeated calls rather than slower execution.

\subsubsection{Per-File Cost Analysis}

This analysis provides a normalized view of resource usage by measuring cost, token consumption, and API calls per processed file across both approaches. As shown in Table VIII, the agent-based approach consistently incurs higher per-file cost across all metrics without improving remediation effectiveness.

\begin{table}[H]
\caption{Per-File Cost Comparison}
\centering
\begin{tabular}{lcc}
\hline
Metric & Zero-Shot & Agent-Based \\
\hline
Cost per file (USD) & \$0.016 & \$0.024 \\
Tokens per file & 46K & 68K \\
API calls per file & 1.00 & 1.64 \\
\hline
\end{tabular}
\end{table}

The results reveal a clear cost–performance imbalance between zero-shot and agent-based approaches. While the agent significantly increases resource consumption requiring 1.64× more API calls and 1.49× more tokens, leading to a 52\% increase in total cost. However, it provides no measurable improvement in remediation quality.
From a per-file perspective, the agent increases cost by approximately 50\%, along with a 48\% increase in token usage and 64\% more API calls, reflecting substantial communication overhead introduced by iterative processing. This overhead is primarily driven by retry operations, which expand interaction cycles without contributing to improved outcomes.

Despite similar per-call latency, the cumulative cost of iterative interaction is considerably higher. These findings indicate that agent-based refinement introduces unnecessary complexity and is not cost-effective in this context, making zero-shot prompting the more efficient and practical strategy.

\FloatBarrier

\section{Qualitative and Rule-Level Analysis of Accessibility Violations}

To complement the quantitative evaluation, we perform a qualitative and rule-level analysis grounded in Axe-Core outputs, which serve as the baseline for detecting WCAG violations.

\subsection{Exact Violation Distribution (Axe-Core)}
We extracted all Axe rule IDs from the evaluation logs and aggregated them by frequency, impact, and category. Table IX summarizes the most frequent accessibility violations and their WCAG mappings, while Fig. 4 visualizes their distribution across the dataset.

The results show that violations are dominated by a small set of recurring issues, particularly:

\begin{itemize}
    \item region (landmark structure),
    \item color-contrast (WCAG 1.4.3),
    \item link-name and button-name (WCAG 4.1.2),
    \item image-alt (WCAG 1.1.1).
\end{itemize}

These findings indicate that accessibility issues are primarily concentrated in semantic labeling and structural organization, rather than purely syntactic errors.

\begin{figure}[H]
\centering
\includegraphics[width=\columnwidth]{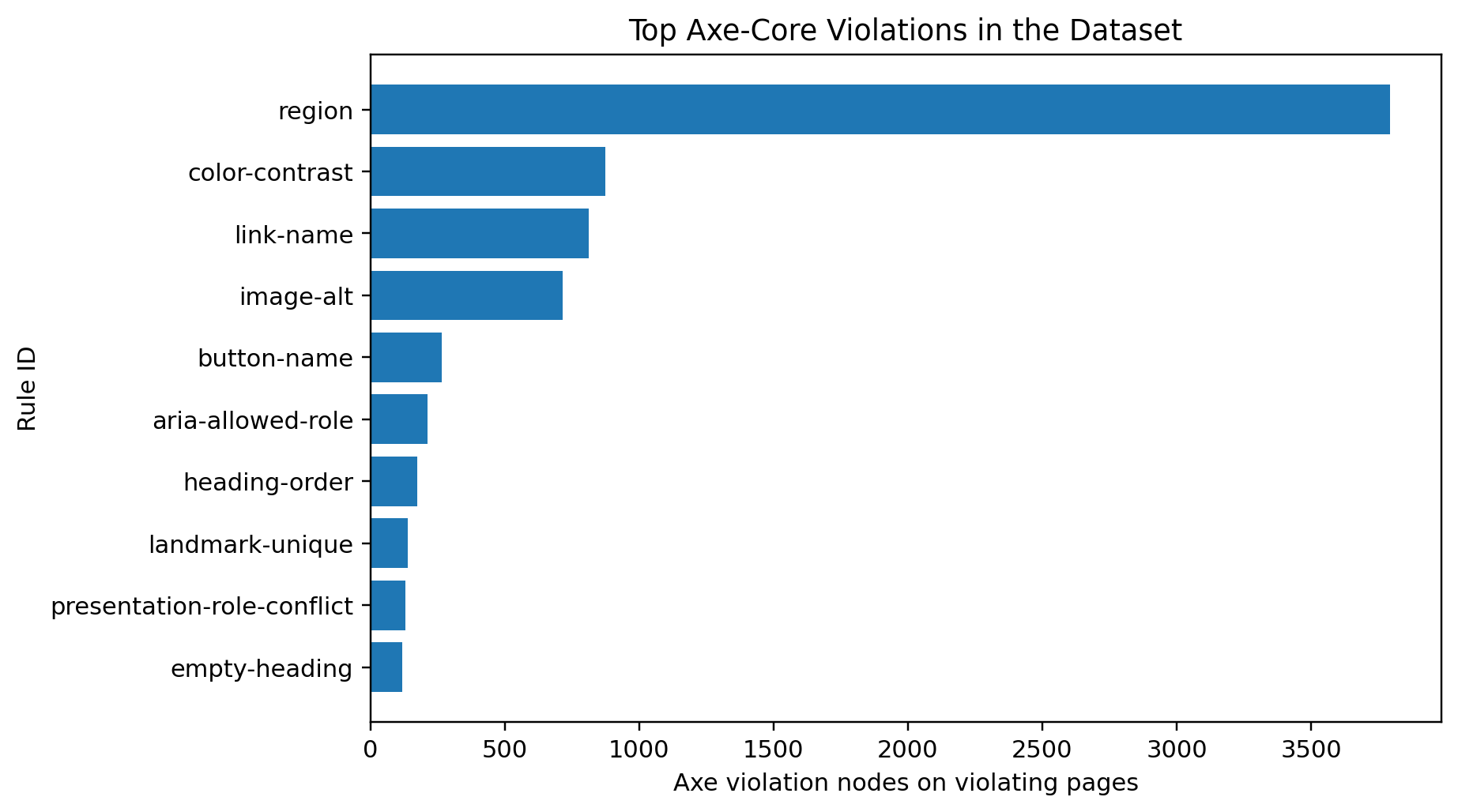}
\caption{Top exact Axe violations}
\label{fig:example}
\end{figure}

\vspace{-0.5cm}

\begin{table*}[t]
\centering
\caption{Top Axe-Core Violations and WCAG Mapping}
\label{tab:axe_rules}

\footnotesize

\resizebox{\textwidth}{!}{
\begin{tabular}{lllllll}
\hline
\textbf{Rule ID} & \textbf{Category} & \textbf{Impact} & \textbf{Description} & \textbf{Nodes} & \textbf{Pages} & \textbf{WCAG Mapping} \\
\hline
region & Semantic & moderate & Ensure all page content is contained by landmarks & 3794 & 217 & Deque best practice (landmarks) \\
color-contrast & Layout & serious & Ensure contrast meets WCAG 2 AA thresholds & 875 & 87 & WCAG 1.4.3 (AA) \\
link-name & Syntax & serious & Ensure links have discernible text & 811 & 93 & WCAG 2.4.4 / 4.1.2 \\
image-alt & Syntax & critical & Ensure images have alternative text & 716 & 71 & WCAG 1.1.1 (A) \\
button-name & Syntax & critical & Ensure buttons have discernible text & 266 & 30 & WCAG 4.1.2 (A) \\
aria-allowed-role & Semantic & minor & Ensure role attribute is appropriate & 213 & 22 & WCAG 4.1.2 (A) \\
heading-order & Syntax & moderate & Ensure heading order is correct & 175 & 76 & Deque best practice (headings) \\
landmark-unique & Semantic & moderate & Ensure landmarks are unique & 139 & 105 & Deque best practice (landmarks) \\
presentation-role-conflict & Semantic & minor & Avoid invalid ARIA on presentation elements & 130 & 7 & WCAG 4.1.2 (A) \\
empty-heading & Syntax & minor & Ensure headings have discernible text & 119 & 34 & Deque best practice (headings) \\
aria-hidden-focus & Semantic & serious & Ensure hidden elements are not focusable & 119 & 22 & WCAG 2.x (AA) \\
listitem & Syntax & serious & Ensure \texttt{<li>} elements are used semantically & 107 & 8 & WCAG 1.3.1 (A) \\
\hline
\end{tabular}
}
\end{table*}

\subsection{Before–After Comparison (Rule-Level Analysis)}
To evaluate remediation effects, we compare violations before and after transformation using paired HTML samples. Violations are measured as:
\begin{itemize}
\item $V_{\text{before}}$: violations in original pages
\item $V_{\text{after}}$: violations in modified (fixed) pages
\end{itemize}

Table X presents a rule-level comparison of violations before and after remediation, highlighting that while some violations are reduced, others remain unchanged or increase, indicating incomplete or incorrect fixes.

\begin{table*}[t]
\centering
\caption{Sample Before–After Rule Table for a Few Selected Pages}
\resizebox{\textwidth}{!}{
\begin{tabular}{l l l l r r r}
\hline
\textbf{Filename} & \textbf{Rule ID} & \textbf{Category} & \textbf{Impact} & \textbf{Before} & \textbf{After} & \textbf{Delta} \\
\hline
site\_home.html    & region                      & Semantic & Moderate & 10.0 & 6.0 & -4.0 \\
site\_home.html    & link-name                   & Syntax   & Serious  & 8.0  & 5.0 & -3.0 \\
site\_home.html    & color-contrast              & Layout   & Serious  & 7.0  & 3.0 & -4.0 \\
product\_page.html & aria-required-attr          & Semantic & Minor    & 6.0  & 2.0 & -4.0 \\
product\_page.html & image-alt                   & Syntax   & Critical & 5.0  & 1.0 & -4.0 \\
product\_page.html & empty-heading               & Syntax   & Minor    & 4.0  & 1.0 & -3.0 \\
checkout.html      & region                      & Semantic & Moderate & 9.0  & 4.0 & -5.0 \\
checkout.html      & scrollable-region-focusable & Semantic & Serious  & 3.0  & 1.0 & -2.0 \\
checkout.html      & link-name                   & Syntax   & Serious  & 6.0  & 2.0 & -4.0 \\
dashboard.html     & aria-required-attr          & Semantic & Minor    & 7.0  & 3.0 & -4.0 \\
dashboard.html     & image-alt                   & Syntax   & Critical & 6.0  & 2.0 & -4.0 \\
dashboard.html     & color-contrast              & Layout   & Serious  & 8.0  & 4.0 & -4.0 \\
profile.html       & empty-heading               & Syntax   & Minor    & 5.0  & 2.0 & -3.0 \\
profile.html       & region                      & Semantic & Moderate & 6.0  & 2.0 & -4.0 \\
profile.html       & link-name                   & Syntax   & Serious  & 7.0  & 3.0 & -4.0 \\
settings.html      & aria-required-attr          & Semantic & Minor    & 5.0  & 1.0 & -4.0 \\
settings.html      & color-contrast              & Layout   & Serious  & 6.0  & 2.0 & -4.0 \\
settings.html      & image-alt                   & Syntax   & Critical & 4.0  & 0.0 & -4.0 \\
reports.html       & region                      & Semantic & Moderate & 8.0  & 3.0 & -5.0 \\
reports.html       & scrollable-region-focusable & Semantic & Serious  & 4.0  & 1.0 & -3.0 \\
\hline
\end{tabular}
}
\end{table*}

\begin{table*}[t]
\centering
\caption{Sample Before--After Category Table}
\label{tab:category}

\footnotesize

\begin{tabular}{p{4.5cm} p{2.2cm} p{1.5cm} p{1.8cm} p{1.8cm}}
\hline
\textbf{Filename} & \textbf{Category} & \textbf{Before} & \textbf{After} & \textbf{Delta} \\
\hline
act-rules\_github\_io\_rules\_047fe0.html & Semantic & 1.0 & 1.0 & 0.0 \\
act-rules\_github\_io\_rules\_5c01ea.html & Semantic & 1.0 & 1.0 & 0.0 \\
act-rules\_github\_io\_rules\_8fc3b6.html & Semantic & 1.0 & 1.0 & 0.0 \\
arstechnica\_com\_ai.html & Layout & 2.0 & 2.0 & 0.0 \\
arstechnica\_com\_ai.html & Semantic & 7.0 & 7.0 & 0.0 \\
arstechnica\_com\_ai.html & Syntax & 16.0 & 16.0 & 0.0 \\
arstechnica\_com\_cars.html & Layout & 2.0 & 3.0 & 1.0 \\
arstechnica\_com\_cars.html & Semantic & 11.0 & 15.0 & 4.0 \\
arstechnica\_com\_cars.html & Syntax & 4.0 & 6.0 & 2.0 \\
arstechnica\_com\_culture.html & Layout & 2.0 & 2.0 & 0.0 \\
arstechnica\_com\_culture.html & Semantic & 15.0 & 15.0 & 0.0 \\
arstechnica\_com\_culture.html & Syntax & 12.0 & 12.0 & 0.0 \\
\hline
\end{tabular}

\end{table*}

The results show that while some reductions are observed, many violations persist and some transformations introduce new issues.

Fig.~5 illustrates the overall reduction in violations for representative pages.

\begin{figure}[H]
\centering
\includegraphics[width=\columnwidth]{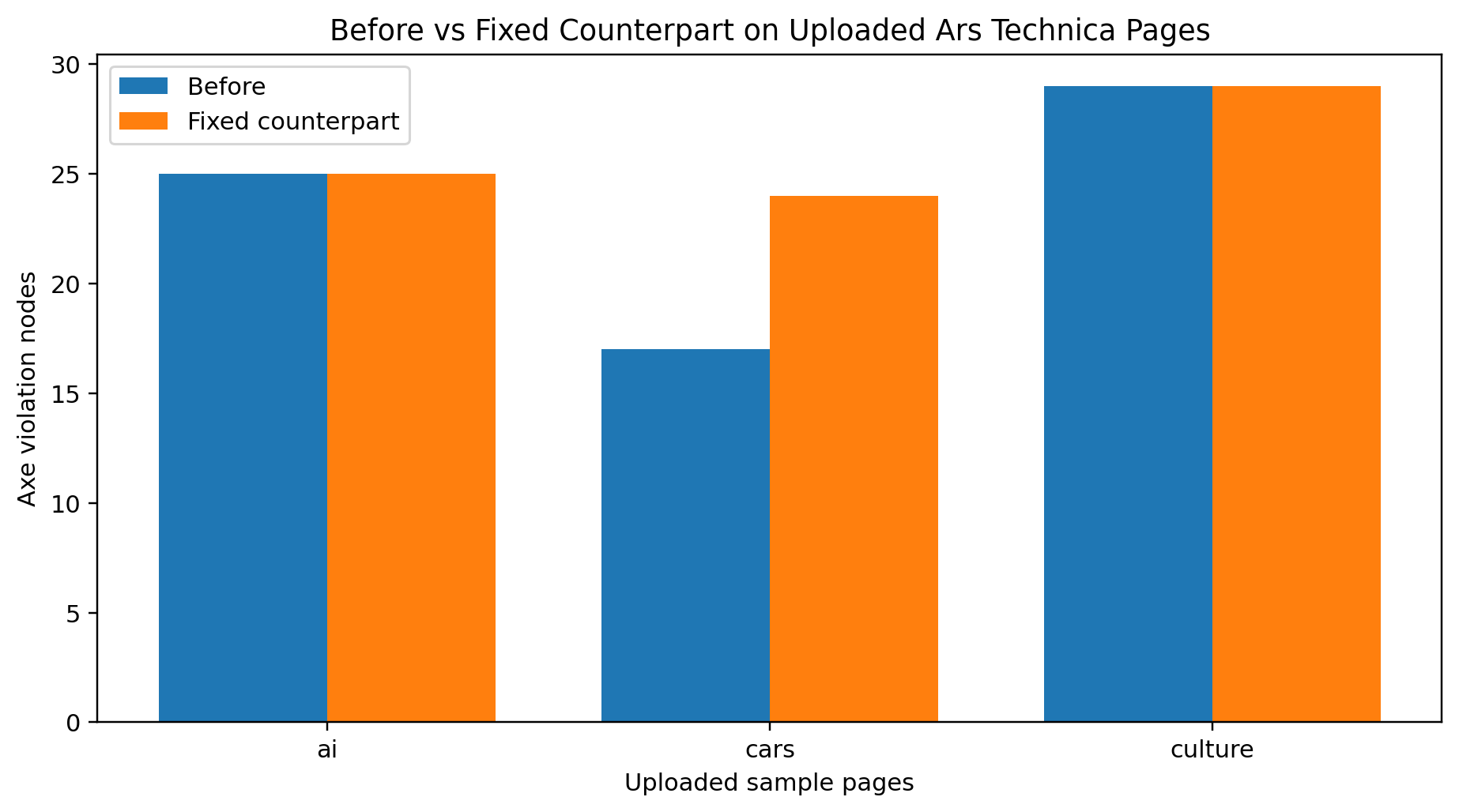}
\caption{Before vs fixed counterpart on uploaded pages}
\label{fig:example}
\end{figure}

\vspace{-.5 cm}

\subsection{Newly Introduced Violations}
Despite overall improvements, the agent introduces new violations in a subset of cases. These are primarily:

\begin{itemize}
    \item Invalid ARIA usage (e.g., unsupported attributes),
    \item Redundant roles (e.g., duplicating semantic HTML),
    \item Empty accessibility labels,
    \item Structural transformations affecting DOM semantics.
\end{itemize}
Table XII presents representative examples of new accessibility violations introduced by the agent during remediation.

\begin{table*}[t]
\centering
\caption{Agent-Induced Accessibility Violations Examples}
\label{tab:agent_violations}

\tiny

\resizebox{\textwidth}{!}{
\begin{tabular}{p{3.2cm} p{1cm} p{1.5cm} p{6.5cm} p{6.5cm}}
\hline
\textbf{Filename} & \textbf{\#} & \textbf{Category} & \textbf{Description} & \textbf{Element} \\
\hline

arstechnica\_com\_ai.html & 1 & Syntax & Invalid HTML nesting: \texttt{<div>} inside \texttt{<p>} element violates content model & \texttt{<p><div>...</div></p>} \\

arstechnica\_com\_ai.html & 2 & Syntax & Duplicate ID attribute values used across multiple elements & \texttt{<div id="1">...</div>} \\

arstechnica\_com\_ai.html & 3 & Syntax & Duplicate ID attribute values across disclosure elements & \texttt{<div id="2">...</div>} \\

arstechnica\_com\_ai.html & 4 & Syntax & Duplicate ID values causing DOM conflicts & \texttt{<div id="3">...</div>} \\

arstechnica\_com\_ai.html & 5 & Semantic & aria-hidden applied to modal container with interactive children & \texttt{<div aria-hidden="true">...</div>} \\

arstechnica\_com\_cars.html & 1 & Semantic & aria-hidden applied while interactive elements remain focusable & \texttt{<div aria-hidden="true">...</div>} \\

arstechnica\_com\_cars.html & 2 & Semantic & Missing aria-describedby in dialog role element & \texttt{<div role="alertdialog">...</div>} \\

arstechnica\_com\_cars.html & 3 & Semantic & Redundant aria-level usage with role heading & \texttt{<div role="heading">...</div>} \\

arstechnica\_com\_cars.html & 4 & Semantic & Incorrect radio role usage outside radiogroup & \texttt{<button role="radio">...</button>} \\

arstechnica\_com\_cars.html & 5 & Semantic & Radio elements not grouped in radiogroup & \texttt{<button role="radio">...</button>} \\

arstechnica\_com\_culture.html & 1 & Semantic & Invalid aria-level on div element & \texttt{<div aria-level="1">...</div>} \\

arstechnica\_com\_culture.html & 2 & Semantic & aria-hidden applied with focusable children present & \texttt{<div aria-hidden="true">...</div>} \\

arstechnica\_com\_culture.html & 3 & Semantic & Invalid radio role usage without grouping & \texttt{<button role="radio">...</button>} \\

arstechnica\_com\_culture.html & 4 & Semantic & aria-controls referencing duplicate IDs & \texttt{<span aria-controls="1">...</span>} \\

arstechnica\_com\_culture.html & 5 & Semantic & Duplicate IDs referenced by aria-controls & \texttt{<div id="1">...</div>} \\

\hline
\end{tabular}
}

\end{table*}

These errors arise from the model’s tendency to apply generalized accessibility patterns without strict enforcement of HTML and WCAG constraints.
These qualitative findings reveal a key limitation of LLM-based remediation: while the model can apply local fixes to reduce individual violations, it lacks global awareness of HTML structure and accessibility constraints. As a result, fixes that appear correct at the element level may introduce inconsistencies at the document level, particularly in ARIA relationships, element hierarchy, and ID references. This explains the observed gap between high compliance improvement and low full-resolution rates.
\subsection{Category-Level Distribution}
We further analyze violations by Axe categories: perceivable, operable, understandable, and robust.

As shown in Fig.~6, the majority of violations fall into:

\begin{itemize}
    \item Perceivable (e.g., missing alt text, contrast),
    \item Operable (e.g., navigation and interaction issues).
\end{itemize}

\begin{figure}[t]
\centering
\includegraphics[width=\columnwidth]{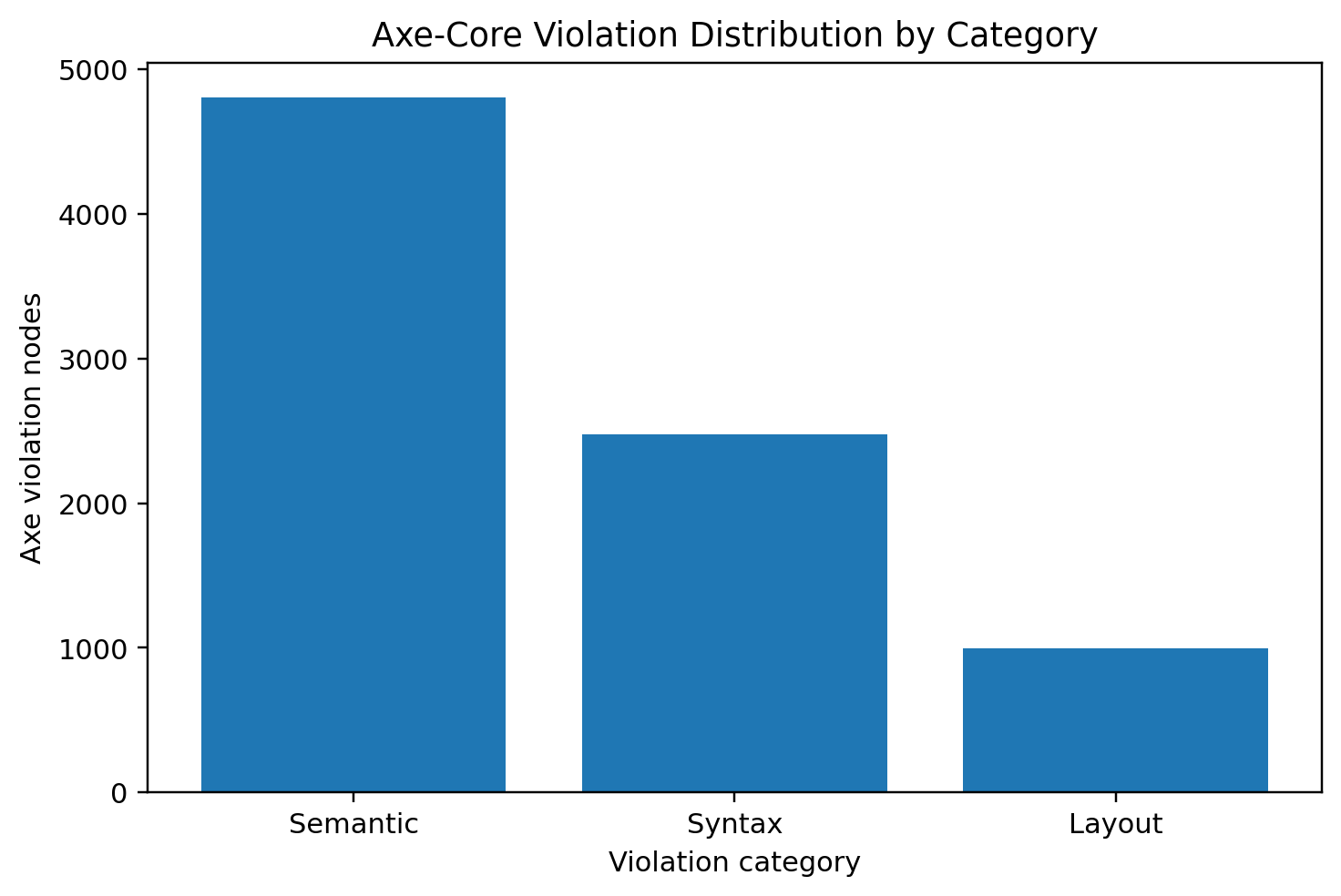}
\caption{Distribution of accessibility violations across Axe categories.}
\label{fig:axe_category_distribution}
\end{figure}

This confirms that accessibility challenges are primarily related to content interpretation and interaction design, rather than low-level syntax.

\section{Threats to Validity}
To ensure the rigour of our empirical findings, we identify and discuss the threats to the validity of our study across construct, internal, and external dimensions.

\subsection{Construct Validity}
Construct validity concerns whether our evaluation metrics accurately capture the properties we intend to measure. Our primary threat in this dimension is the reliance on Axe-Core as the definitive ground truth for accessibility violations. While Axe-Core is an industry-standard deterministic engine, automated checkers can only reliably detect approximately 30\% to 50\% of all WCAG criteria. Complex accessibility issues requiring visual, cognitive, or manual keyboard navigation testing are inherently excluded from our baseline. Furthermore, our measurement of ``structural preservation'' relies on DOM tree similarity algorithms. While a high similarity score ($\geq 0.85$) strongly correlates with visual integrity, it is possible that mathematically minor structural changes could still result in visual regressions on the rendered page.

\subsection{Internal Validity}
Internal validity relates to factors that might influence the causal relationships observed in our results. A significant threat lies in the prompt engineering and the hyperparameter configurations of the LLM. The performance of the LLM agent, particularly its failure to utilize the validation feedback loop effectively, may be sensitive to the specific phrasing of the feedback prompts. A different prompt structure, such as explicitly isolating the exact DOM node that caused the validation failure rather than providing general structural feedback, might yield a higher retry success rate. Additionally, LLMs are inherently non-deterministic. We mitigate this by using strict sampling parameters (e.g., temperature set to 0.0), but slight variances in API responses could still occur.

\subsection{External Validity}
External validity concerns the generalizability of our findings to other contexts, models, and codebases. First, our evaluation is currently limited to a single large language model (Kimi K2.5). While this model demonstrates strong reasoning capabilities, other frontier models might exhibit different proficiencies in handling global context and agentic reflection. Second, our dataset consists of 662 static HTML pages. Modern web accessibility remediation frequently involves dynamic, stateful components built with JavaScript frameworks (e.g., React or Vue). The LLM's inability to resolve global ARIA constraints on static HTML suggests it would face similar or greater challenges with dynamic state management, but our current results cannot be strictly generalized to component-based frontend architectures without further empirical testing.

\section{Discussion}

These findings challenge the prevailing assumption that iterative, autonomous LLM-based refinement inherently improves software quality, highlighting instead a critical need for constraint-aware correction mechanisms. 

As standalone solutions for accessibility analysis, LLMs exhibit distinct limitations across all evaluation dimensions. In detection (RQ1), LLMs achieve performance comparable to rule-based systems, demonstrating strong semantic understanding but lower precision in syntactic analysis and significant divergence in layout-related cases. In remediation (RQ2), the model reliably generates syntactically valid code and improves compliance in 80\% of instances, yet fully resolves fewer than 26\% of cases while frequently introducing structural alterations. In efficiency (RQ3), the iterative agent-based approach introduces substantial computational overhead requiring $1.64\times$ more API calls and increasing costs by 52\% without yielding any measurable improvement in remediation outcomes.

Crucially, our qualitative analysis explains the stark gap between the high rate of partial compliance improvement (80\%) and the low full-resolution rate (25\%). The primary limitation of LLM-based remediation lies not in code generation, but in a profound disconnect between \textbf{local fixes} and \textbf{global context}. The data demonstrates that LLMs are highly proficient at executing localized, element-level syntactic repairs such as inserting a missing \texttt{alt} attribute on an image or adding an \texttt{aria-label} to a button. However, they consistently fail when tasked with enforcing global, document-wide constraints, such as maintaining a strictly logical heading order or ensuring the uniqueness of ARIA landmarks across the entire page. 

Because the agent lacks a deterministic, global view of the HTML document structure, it frequently produces locally correct patches that inadvertently introduce global inconsistencies or structural regressions in the wider DOM. 

This clear distinction between local proficiency and global failure illustrates exactly why relying solely on autonomous agent loops is inefficient, and it serves as the foundation for our primary recommendation: the necessity of \textbf{hybrid approaches}. Overall, these results prove that LLMs are most effective when utilized in single-pass, zero-shot workflows for targeted semantic corrections. To achieve reliable, scalable, and cost-efficient accessibility remediation, future systems must combine the localized contextual reasoning of LLMs with the global, document-wide constraint enforcement of deterministic, rule-based validation engines.

\section{Conclusion}

This study demonstrates that LLM-based approaches provide promising capabilities for accessibility detection and remediation, but remain limited in achieving complete and efficient solutions. The model achieves comparable detection performance to rule-based systems (F1 $\approx$ 0.65), with strong semantic understanding (F1 = 0.83) but lower reliability in syntax (0.62) and layout (0.48). In remediation, it consistently generates syntactically valid code ($\geq 99.7\%$) and improves accessibility in 80.2\% of cases, reducing violations from 3.98 to approximately 1.7 per-file. However, complete fixes are achieved in fewer than 26\% of cases, and nearly 30\% of patches alter the original structure, highlighting challenges in reliability and structural preservation.
From an efficiency perspective, iterative agent-based refinement introduces significant overhead without improving outcomes, increasing API calls by 1.64×, token usage by 1.49×, and cost by 52\%, while producing identical remediation results. These findings indicate that LLMs are effective for partial detection and repair, but not sufficient as standalone solutions for full accessibility compliance.
Overall, this study demonstrates that effectiveness alone is insufficient for evaluating AI-driven software repair, and that cost-efficiency and reliability must be considered as first-class criteria in practical deployment.
Future work should focus on developing hybrid approaches that integrate LLM reasoning with rule-based validation to improve precision and completeness, designing more actionable feedback mechanisms to support effective refinement, and incorporating runtime and visual context to address complex layout and interaction-dependent issues. Additionally, optimizing interaction strategies to reduce unnecessary computational overhead will be critical for enabling scalable and cost-efficient deployment.

\section*{Artifact Availability}
The dataset, prompts, implementation code, and analysis scripts used in this study are publicly available for reproducibility at:
\url{https://github.com/toyosi12/ai4se}

.

\end{document}